%% file: main.tex
\documentclass[10.75pt, a4paper]{article}
\input{initialisation.tex}
\usepackage[framemethod=tikz]{mdframed}
\usepackage{lipsum}
\usepackage{dirtytalk}
\usepackage{tcolorbox}
\usepackage[table]{xcolor}
\usepackage{caption}
\newtcolorbox{mybox}[1]{colback=green!6!white,colframe=black!75!black,fonttitle=\bfseries,title=#1}
\newtcolorbox{mybox2}{colback=red!5!white,colframe=red!75!black}

\usepackage{pifont}

\def\be{\begin{equation}}
\def\ee{\end{equation}}

\usepackage{soul,xcolor}
\setstcolor{red}

\usepackage{xcolor,hyperref}
\hypersetup{
   colorlinks,
   linkcolor={blue!50!black},
   citecolor={blue!50!black},
   urlcolor={blue!80!black}
} 

\definecolor{mycolor}{rgb}{0.122, 0.435, 0.698}

\title{Cage Breaking Far from Equilibrium}

\author[1]{Jared Popowski\footnote{j.s.popowski@uva.nl, ORCID: 0009-0001-7507-5651}}
\author[1,2]{Nico Schramma}
\author[3]{Edan Lerner}
\author[1, 4]{Maziyar Jalaal\footnote{m.jalaal@uva.nl, ORCID: 0000-0002-5654-8505}}

\affil[1]{Van der Waals-Zeeman Institute, Institute of Physics, \protect\\
University of Amsterdam, Science Park 904, Amsterdam 1098XH, The Netherlands}
\affil[2]{Department of Medical Biochemistry,
\protect\\
Amsterdam UMC University of Amsterdam, Meibergdreef 9, Amsterdam 1105AZ, The Netherlands}
\affil[3]{Institute of Theoretical Physics, University of Amsterdam, \protect\\
Science Park 904, Amsterdam 1098XH, The Netherlands}
\affil[4]{Department of Applied Mathematics and Theoretical Physics, \protect\\
University of Cambridge, Wilberforce Road, Cambridge CB3 0WA, United Kingdom}

\begin{document}
\input{colors_define.tex}
\begingroup
\sffamily
\date{}
\maketitle
\endgroup

\begin{abstract}
Active matter can flow and yield under conditions where passive matter jams and slows down, as self-propulsion significantly modulates particle escape from local cages. How activity microscopically reshapes the caging environment to produce this effect, however, remains poorly understood. Here we study a minimal active-matter model of cage breaking: three distinguishable self-propelling disks under circular confinement. This simple setting allows us to construct an entropic landscape for rearrangements and to compare it exactly with its equilibrium counterpart. At low activity the landscape is effectively bistable, whereas at high activity it develops additional metastable basins associated with frustrated clusters at the boundary. We quantify the system's departure from equilibrium and show that cage breaking is fastest when the persistence length matches the particle radius, linking a geometric microscopic scale to the enhanced dynamics of active glasses. Extending the landscape to two dimensions reveals circulating probability currents, and a Markov-state description shows that detailed balance is broken both in the continuous landscape dynamics and in the coarse-grained transitions between entropic basins. Our results provide a minimal microscopic framework for understanding how activity reshapes caging, relaxation, and irreversibility in dense nonequilibrium matter.

\textbf{keywords: active matter $|$ entropic landscape $|$ cage breaking $|$ non-equilibrium thermodynamics $|$ metastability} 

\end{abstract}

\section*{Introduction}

Many natural and synthetic systems operate at high densities, where motion is strongly constrained by neighboring particles. Cage breaking (informally, particles swapping positions with their neighbors) is the elementary rearrangement event in dense, slowly-relaxing particle systems, ranging from glasses to vibrated granular materials and biological collectives~\cite{Fielding2000AgingMaterials, Reis2007CagingFluid,Nishikawa2024CollectiveModel,Debets2021CageMatter,Schramma2023ChloroplastsConditions}. This process can be understood as escape from a metastable basin, a problem that plays a central role in many contexts such as protein conformation, memory in Landauer-limited computation, electron transport in semiconductors, and relaxation of amorphous solids \cite{Onuchic1997TheoryPerspective,Haneef2020ChargeDevices, Pratt2026MetastablePrimer, Sollich1997RheologyMaterials, Schuler2005ExperimentalRates, Militaru2021EscapePotentials}.
Analysis of equilibrium escape dynamics was placed on a rigorous foundation by Kramers, who used the separation of timescales between intra-basin fluctuations and inter-basin transitions to define a single thermal rate constant \cite{Kramers1940BrownianReactions}. In his work and ensuing developments, varying degrees of implicit or explicit reference to the concept of the potential energy landscape of the system are made \cite{Hanggi1990Reaction-rateKramers, Stillinger1995AFormation,Wales2006PotentialLandscapes}. However, the exact computation of such a landscape becomes intractable for high dimensional systems \cite{Goldstein1969ViscousPicture}. As a result, there is extensive work on system dynamics on dimensionally-reduced landscapes in data-rich fields such as chemical kinetics, genetics, or computer science \cite{Kauzmann1948TheTemperatures, Waddington1957TheBiology, Stillinger1984PackingSolids, Stillinger1995AFormation, Hamprecht2001AVisualization, Schweizer2003EntropicSuspensions, Schweizer2004ActivatedLiquids, Charbonneau2014FractalGlasses, Lopez-Alamilla2018ReconstructingPotentials, Jacobs2018AccurateLandscape, Lee2025TopologicalMatter}.

Basin escape in active matter systems, in which constituent particles generate persistent self-propulsion forces, has received growing attention, motivated by its relevance to biological processes ranging from cells squeezing through constrictions \cite{Bruckner2019StochasticSystems} to bacteria escaping pores in disordered media \cite{Bhattacharjee2019BacterialMedia, Bhattacharjee2019ConfinementMedia}. A central question is how self-propulsion modifies escape rates beyond the thermal Kramers framework, and recent experimental and theoretical works have investigated transition rates of active and nonequilibrium matter in the presence of prescribed potentials \cite{Schuler2005ExperimentalRates, Woillez2019ActivatedState, Woillez2020ActiveModel, Wexler2020DynamicsTrap, Dago2021InformationOscillator, Militaru2021EscapePotentials, Basu2025ErgodicityPotential}. However, in many dense systems the relevant landscape that particles navigate emerges from their mutual interactions and local packing geometry as opposed to being externally imposed. For systems with hard-core interactions the distinction is especially clear: particles must thread through dynamic, geometrically constrained passages in their exploration of a configuration space with a flat energy landscape, and the Helmholtz free energy is entropically dominated, $F(\mathbf{x})=-TS(\mathbf{x})$. While thermodynamic concepts such as the free energy or temperature are often ill-defined for nonequilibrium systems, 
the (negative) entropic landscape $-S(\mathbf{x})$ remains valid and useful for characterizing transitions between metastable states in active matter, as it expresses the likelihood of occupying different regions of configuration space, reduces to the free energy at equilibrium when potential energy is negligible, and its barriers directly set the timescale for structural rearrangement.

\begin{figure}[!tbph]
    \centering
    \includegraphics[width=0.85\textwidth]{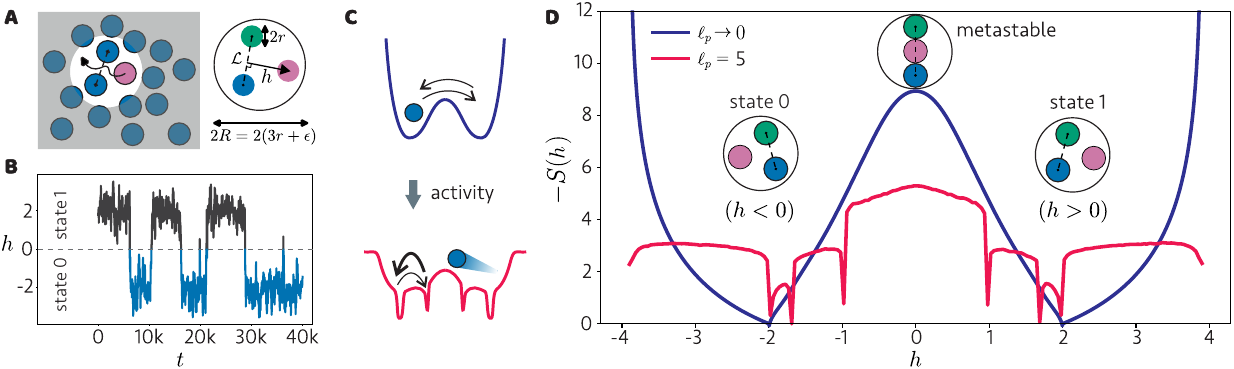}
    \caption{\textbf{Two-state model system for cage breaking becomes multistable with activity.} \textbf{A.} Schematic of the mapping from bulk systems to our minimal model, and definitions of geometric parameters. \textbf{B.} Dynamics in $h$--space are intermittent for tight confinement ($\epsilon=0.01, \ell_p=0.2$), with the system spending long times between sign changes of $h$. \textbf{C.} Illustration of the effect of activity on the entropic landscape, with balanced transitions on a bistable landscape near equilibrium and unbalanced transitions on a multistable landscape far from equilibrium. \textbf{D.} Entropic landscape for the passive and active systems under tight confinement, $\epsilon=0.077$. The $\ell_p \to 0$ curve is the analytic result for a passive Brownian hard-disk system at this confinement (Ref. \cite{Hunter2012Free-energyDisks}), while the $\ell_p=5$ curve is simulated.}
    \label{fig1:system-explanation}
\end{figure}

Here, we study cage breaking and metastability of three self-propelled hard disks under circular confinement, a minimal model for caging in dense active systems (see Fig.~\ref{fig1:system-explanation}A). Projecting the dynamics onto a single reaction coordinate $h$, we show that an entropic landscape $-S(h)$ can be defined and is informative, discovering a transition from two-state to multi-state landscapes as activity increases. The additional states correspond to geometric arrangements of disks in contact on the boundary, whose transitional dynamics we study via a Markov state model \cite{Das2022DrivenEnvironments, Mungan2025Self-OrganizationMatter} that we find breaks detailed balance at high activity strength. We also discover that cage breaking in this minimal model is optimized when the persistence length of the motion matches the particle radius, and develop a novel method to measure the distance from equilibrium of the entropic landscape by means of the Wasserstein-1 distance from optimal transport. Based on these results, we argue that the construction of low-dimensional entropic landscapes from observational data is useful in a wider range of nonequilibrium contexts as a bridge to the tools of thermodynamics.

Our configuration follows previous minimal models of hard-disk systems in thermal equilibrium, where analytic calculation of thermodynamic properties such as the free energy landscape is tractable~\cite{Speedy1994TwoBox,Bowles1999FiveBox, Hunter2012Free-energyDisks, Du2016EnergyModel, Weeks2020VisualizingDisks}. We initialize simulations by placing three distinguishable quasihard disks of effective radius $r_{\mathrm{eff}}$ in random non-overlapping positions within a fixed quasihard-walled circular confinement of radius $R=3\,r_{\mathrm{eff}}+\epsilon$, where the confinement parameter $\epsilon$ controls the system density~\cite{Hunter2012Free-energyDisks} (see Fig.~\ref{fig1:system-explanation}A). Each particle $i$ is described by the following athermal overdamped Langevin equation \cite{Debets2021CageMatter, Fily2012AthermalAlignment}: 
\be
    \dot{\mathbf{r}}_{i}=\zeta^{-1}(\mathbf{F}_i + \mathbf{f}_i), \label{eq: eq-of-motion}
\ee
where $\mathbf{r}_i$ denotes the position of particle $i$, $\zeta$ is the damping coefficient, and $\mathbf{F}_i$ and $\mathbf{f}_i$ are the interaction and self-propulsion forces acting on particle $i$, respectively modeled with a Weeks-Chandler-Andersen \cite{Weeks1971RoleLiquids} (WCA) potential with interaction energy $E_{\mathrm{int}}$ and an active Brownian particle (ABP) self-propulsion force with magnitude $f$ and persistence time $\tau_p$ (Supplementary Material). The circular confinement is endowed with the same WCA repulsive potential as the disks. Note that the only source of noise in the particle dynamics is due to the self-propulsion force. We study overdamped systems that lack momentum conservation as these have been shown to be good approximations to the dynamics of systems as diverse as granular materials or living organisms on a substrate \cite{Fily2012AthermalAlignment}. We use the persistence length $\ell_p=\zeta^{-1}f\tau_p$ of the active Brownian particles as a control parameter for activity strength. All results are presented in reduced units where $r_{\mathrm{eff}}, E_{\mathrm{int}},$ and $t_{\mathrm{red}}\equiv\zeta r_{\mathrm{eff}}^2/E_{\mathrm{int}}$ represent the units of length, energy, and time, respectively.

The state space of our active three-disk system is nine-dimensional and is completely described by all allowable (\emph{i.e.} non-overlapping) combinations of the disk positions and orientations of their self propulsion vectors $(\vec{r}_1, \vec{r}_2, \vec{r}_3, \theta_1, \theta_2, \theta_3)$. However, cage-breaking events in this model correspond to particle rearrangements that are fully captured by the relative position of one disk with respect to the other two, so in the spirit of Ref. \cite{Hunter2012Free-energyDisks} we collapse this space into a single dimension. We designate one disk as the tracked disk and define $h$ as its signed perpendicular distance to the line connecting the other two, where the sign encodes which side of that line it occupies (Fig \ref{fig1:system-explanation}A). Cage-breaking events are then equivalent to the times when $h$ changes sign (Fig \ref{fig1:system-explanation}B). This projection of the 9D state space onto a 1D representative coordinate is not unique and many other variables could be chosen that capture the fundamental caging dynamic in this model \cite{Hunter2012Free-energyDisks}, but we utilize $h$ for its simple geometric interpretation as the height of the triangle formed by the disk centers.

For each simulation timestep, the macrostate $h$ is computed from the disk positions, resulting at the end of simulations in a histogram $n(h)$ that represents the system's estimated multiplicity. The negative entropy relative to the ground state of the system (the most frequently observed $h$-value $h_0$) is then computed as
\be
    -S(h) = -\log[n(h)/n(h_0)]. \label{eq: entropic_landscape_def}
\ee

\begin{figure}[!tbph]
    \centering
    \includegraphics[width=0.47\textwidth]{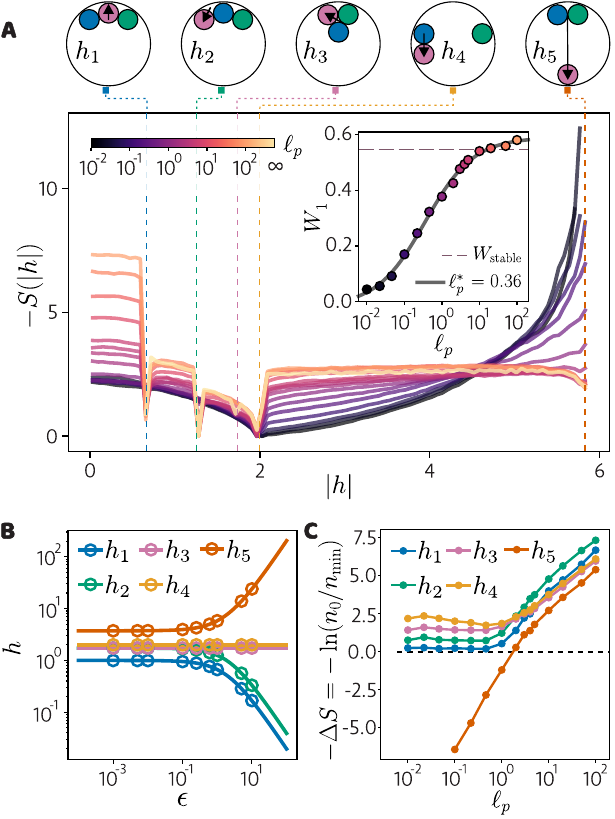}
    \caption{\textbf{Persistent particles reshape their entropic landscape through clustering on the confinement boundary.} \textbf{A.} Evolution of the entropic landscape as a function of the persistence time, $\epsilon=1$. Here we plot each landscape relative to its deepest minimum, and as a function of $|h|$ since the landscape is symmetric with respect to $h$. The landscape arising from sampling of stable configurations on the boundary is labeled $\ell_p=\infty$. Above: typical clustered configurations of the particles that cause the minima in the entropic landscape through activity frustration, with a black arrow showing their h values. Inset: the Wasserstein-1 distance $W_1$ between the simulated landscapes and the analytic distribution of the system at equilibrium (Ref. \cite{Hunter2012Free-energyDisks}) monotonically increases with $\ell_p$ and quantifies the landscape's distance from equilibrium. Points are data from simulated landscapes, black curve is a fit to a sigmoid function. $W_{\mathrm{stable}}$ is the Wasserstein-1 distance of the stable configurations distribution at $\ell_p=\infty$. \textbf{B.} The locations of the system's entropic landscape minima as confinement strength $\epsilon$ varies, as predicted by geometry (solid curves) and extracted from stable configuration simulations with an extrema detection algorithm (open markers). \textbf{C.} Entropic depths of the points on the simulated landscapes from panel A at each minima location relative to the entropic barrier at $h=0$, as a function of persistence time. The crossover where the depth at $h_2$ becomes maximal indicates that it transitions to being the most sampled minima at high activity for this confinement size.}
    \label{fig2:entropic-landscape}
\end{figure}

The entropic landscape $-S(h)$ is a symmetric function of $h$ (see Fig. \ref{fig1:system-explanation}D), as Eq. (\ref{eq: eq-of-motion}) is invariant under the particle exchange $i \leftrightarrow j$ that is necessary for $h$ to change signs. Therefore to study the origin of the additional minima in the entropic landscape, we combine statistics from both signs of $h$, folding the landscape onto $-S(|h|)$ as shown in Fig. \ref{fig2:entropic-landscape}A. These minima correspond to $|h|$ values that the active system oversamples with respect to the equilibrium distribution via geometric arrangements of the disks in contact on the boundary. Due to their persistent motion, dry active systems generically accumulate on confinement boundaries and corners \cite{Galajda2007ABacteria, Li2009AccumulationMotion, Kaiser2012HowParticles, Bechinger2016ActiveEnvironments}, and we observe the same for our model. Particles driven into the boundary or each other can become transiently trapped in a frustrated force balance, remaining there until they reorient after a time $\mathcal{O}(\tau_p)$. Unlike the equilibrium system which has a wide minimum for $|h|>2$, the minima arising from activity are localized dips as they correspond to particular geometric configurations, and we label them $h_i$ in order of increasing $|h|$. The apparent width of these minima in the entropic landscape is due to finite bin widths in the histogram $n(h)$.

How far from equilibrium are the dynamics at some nonzero value of $\ell_p$? To quantify the distance from equilibrium, we compute the Wasserstein-1 distance (aka the earth mover's distance) between the system's equilibrium and nonequilibrium probability distributions in $h$--space. Denoting the probability distributions by $u(|h|),v(|h|)$ and their cumulative distribution functions by $U(|h|), V(|h|)$ respectively, the Wasserstein-1 distance is
\be
    W_1(u,v)=\int_{0}^{\infty}\big|U-V\big|\,d|h|.
\ee
Given that Ref. \cite{Hunter2012Free-energyDisks} derives the analytic distribution for the equilibrium case, our computation of the Wasserstein-1 distance is exact for our simulations. We find that the data follow a sigmoid function, which we can fit to obtain the turning point $\ell_p^*=0.36$ and saturated Wasserstein-1 distance at infinite persistence, $W_1^{\infty}=0.59$ (see inset of Fig. \ref{fig2:entropic-landscape}A). The particular values of $\ell_p^*$ and $W_1^{\infty}$ depend on the confinement $\epsilon$, see the Supplementary Material for $\epsilon=0.25$. The turning point quantifies a value of $\ell_p$ for which the landscape transitions between near- and far-from-equilibrium behavior. This represents a novel measure for the distance to equilibrium in active matter, which has traditionally studied violation of the equilibrium fluctuation-dissipation theorem or the entropy production rate \cite{Fodor2016HowMatter, Gnesotto2018BrokenReview, OByrne2022TimeMacro, Nartallo-Kaluarachchi2025Coarse-grainingChains}.

To verify that the entropic landscape's minima emerge from active exploration of metastable disk arrangements on the boundary, we accumulate $n(h)$ from samples of stable force-balanced configurations reached by disks with random orientations and infinite persistence time, see Supplementary Material for detail. The resulting entropic landscape ($\ell_p=\infty$ curve in Fig. \ref{fig2:entropic-landscape}A) has the same shape as in the finite-persistence case, with a Wasserstein-1 distance $W_{\mathrm{stable}}=0.55$ close to the saturated fit value of $W_{1}^{\infty}=0.59$. The slight undershoot is likely due to the difference in sampling procedures, as the stable configuration sampling misses dynamical information present in the finite-$\ell_p$ simulations, discussed in more depth in the Supplementary Material. In Fig. \ref{fig2:entropic-landscape}B, we compare extrema detected in the infinite-persistence $n(h)$ against the analytically calculated $h_i(\epsilon)$ for each of the disk arrangements at the top of Fig. \ref{fig2:entropic-landscape}A as $\epsilon$ is varied (see Supplementary Material for $h_i(\epsilon)$ equations). We find that there are precisely five entropic minima in $h$--space whose locations match the values dictated by the geometry of the stable configurations, with two of these minima becoming indistinguishable in the tight confinement limit $\epsilon \to 0$, since $h_2(0)=h_3(0)$.

The entropic well depth of each of the metastable minima relative to the entropic barrier at $h=0$, $-\Delta S=-\ln(n_{0}/n_{\mathrm{min}})$, is plotted in Fig. \ref{fig2:entropic-landscape}C. Near equilibrium $\ell_p \ll \ell_p^{*}$, the system remains effectively bistable and the depths simply reflect the shape of the underlying equilibrium landscape, so the curves are flat. As the system gets further from equilibrium with $\ell_p \geq \ell_p^*$, the depths of each of the minima increase, until there is a crossover in the system's deepest minimum from $h_4$ to $h_2$. Thus our active system spends significantly more time in a clustered arc on the boundary than at the separation of $h=2$ that is entropically favored near thermal equilibrium (compare deepest minima at low and high $\ell_p$ in Fig. \ref{fig2:entropic-landscape}A).

The cage breaking time is computed as the average time between sign changes of $h$, $\tau = \dfrac{1}{N}\sum_i\Delta t_i.$ There are often small fluctuations around $h=0$ that are not true cage breaking transitions, so to avoid biasing $\tau$ to smaller timescales we set a distance threshold of $\Delta h=1$ that the disk must move after crossing $h=0$ before the transition is considered completed \cite{Hunter2012Free-energyDisks}. Mean cage breaking times are not sensitive to this choice \cite{Du2016EnergyModel}.

\begin{figure}[!tbph]
    \centering
    \includegraphics[width=0.85\textwidth]{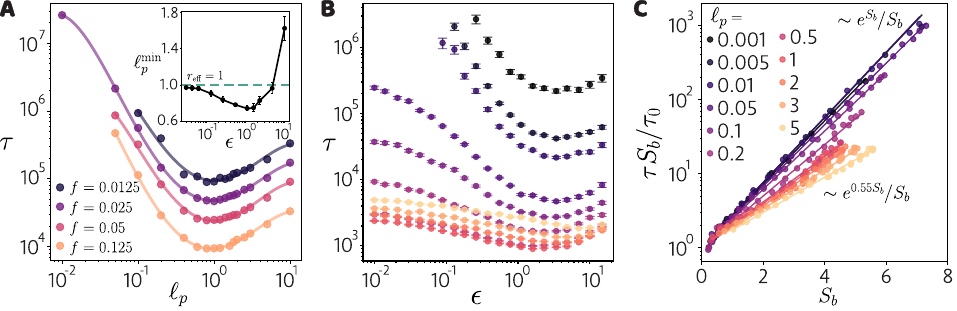}
    \caption{\textbf{Optimal cage breaking time at intermediate activity; caging dependence on confinement.} \textbf{A.} The caging time is nonmonotonic with persistence length, minimized at an intermediate value $\ell_{p}^{\mathrm{min}}$ whose location does not depend on the self-propulsion force $f$. Points are simulated data, curves are spline fits to a quartic polynomial in log-log space. Here $\epsilon=0.062$ and $E_{\mathrm{int}}=1000$. Inset: Values of $\ell_{p}^{\mathrm{min}}$ extracted from fits of data with $f=0.025$, $E_{\mathrm{int}}=1000$ and various values of $\tau_p$ and $\epsilon$ is consistent with the particle's effective radius $r_{\mathrm{eff}}=1$ under tight confinement (small $\epsilon$), where cage breaking is well-defined with a single pathway. \textbf{B. } As the confinement size varies for fixed persistence length, the caging time finds an optimum at $\epsilon > 1$ (exact value depends on the simulated persistence length). Here and in panel C $E_{\mathrm{int}}=40$. Data are colored according to the legend in panel C. \textbf{C. } A modified version of Kramer's law (Eq. \ref{eq: kramers-scaling}) accurately captures caging times over all entropic barrier depths $S_b$ and levels of activity. The slope transitions with increasing $\ell_p$ from the value of 1 expected at equilibrium to smaller values as cage breaking is enhanced by activity. Caging time data is normalized by the prefactor $\tau_0(\ell_p)$ from a fit of Eq. (\ref{eq: kramers-scaling}) to each dataset.}
    \label{fig3:caging-times}
\end{figure}

To accurately compute $\tau$, we run simulations of $\mathcal{O}(10^{9}-10^{10})$ time steps for various persistence lengths $\ell_p$ under tight confinement $\epsilon=0.062$ and steep WCA interaction energy $E_{\mathrm{int}}=1000$ to mimic hard disks (see Fig. \ref{fig3:caging-times}A). Strikingly, we observe a nonmonotonic dependence of $\tau$ with $l_p$, with optimal cage breaking (minimal $\tau$) at intermediate persistence. Fitting the data to a quartic polynomial in log-log space allows us to extract the exact persistence length $\ell_p^{\mathrm{min}}$ that minimizes $\tau$, whose value we find does not depend on $f$. For tight confinement, cage breaking is optimized at the particle's effective hard-disk radius $\ell_p^{\mathrm{min}}=r_{\mathrm{eff}}$ (inset, Fig. \ref{fig3:caging-times}A).

Our system's second control parameter for cage breaking is the free space available for the disks to rearrange, tuned by varying $\epsilon$. We run simulations over many values of $\ell_p$ from $0.001-5$ and $\epsilon$ from $10^{-2}-10^{1}$ with $E_{\mathrm{int}}=40$ to investigate the dependence of $\tau$ on $\epsilon$, see Fig. \ref{fig3:caging-times}B. Similarly to the effect of $\ell_p$, we find that disk rearrangements strongly and nonmonotonically depends on $\epsilon$. Cage breaking time $\tau$ is minimized for an intermediate confinement parameter $\epsilon >1$ whose exact value varies nonmonotonically with $\ell_p$ and is not our focus here, as this level of confinement is already sufficiently weak to allow for multiple different pathways for cage breaking. For even larger values of $\epsilon$ the system enters the ``dilute active gas" limit, where disks must travel substantially farther on average to complete a rearrangement, leading to an increase in $\tau$.

Near equilibrium, the caging time increases dramatically as $\epsilon$ shrinks. In the limit of deep entropic barriers $S_b \gtrsim 7$ (where $S_b$ is defined in the following paragraph) the near-equilibrium data are expected to approach a power law relation of $\tau \sim \epsilon^{-7/2}$ \cite{Hunter2012Free-energyDisks}, but our simulations are in an intermediate regime of barrier depths for which this scaling does not yet hold. Cage breaking is facilitated as the particles become more persistent, decreasing the slope of the curves in Fig. \ref{fig3:caging-times}B until they become nearly flat. This can be understood as the system transitioning from diffusive exploration to ballistic exploration bounded by the confinement geometry, see Supplementary Material.

Finally, in analogy with the definition of the free energy barrier $F_{B}$ in the equilibrium case, we define the entropic barrier to a cage breaking transition $S_b(\ell_p, \epsilon)$ as the difference between the negative entropies of the landscape's deepest minimum and the value at $h=0$. We discover that $\tau$ follows a modified Kramers-like scaling across all depths of entropic barriers, even far from equilibrium,
\be
    \tau = \tau_0 \exp(\beta^{*}S_b)/S_b,\label{eq: kramers-scaling}
\ee
where $\tau_0(\ell_p)$ and $\beta^*(\ell_p)$ are fit parameters. We verify the scaling of Eq. \ref{eq: kramers-scaling} in our system by plotting $\log(\tau S_b)/\tau_0$ in Fig. \ref{fig3:caging-times}C. Note that an equilibrium system in natural units ($k_B=1$) should have $\beta^*=1$, and in the limit of very large $S_b$ this equation indeed approaches the typical Kramers scaling $\tau\sim \exp(S_b)$. We find that the fit slope begins at this equilibrium value of $\beta^{*}=1$ for low persistence simulations and monotonically decreases to $\beta^{*}\approx 0.55$ at $\ell_p = 5$. As $S_b(\ell_p, \epsilon)$ depends on both the level of activity and the degree of confinement, Fig. \ref{fig3:caging-times}C implies that increased persistence always enhances relaxation dynamics for systems that possess the same entropic barrier.

\begin{figure}[!tbph]
    \centering
    \includegraphics[width=0.85\textwidth]{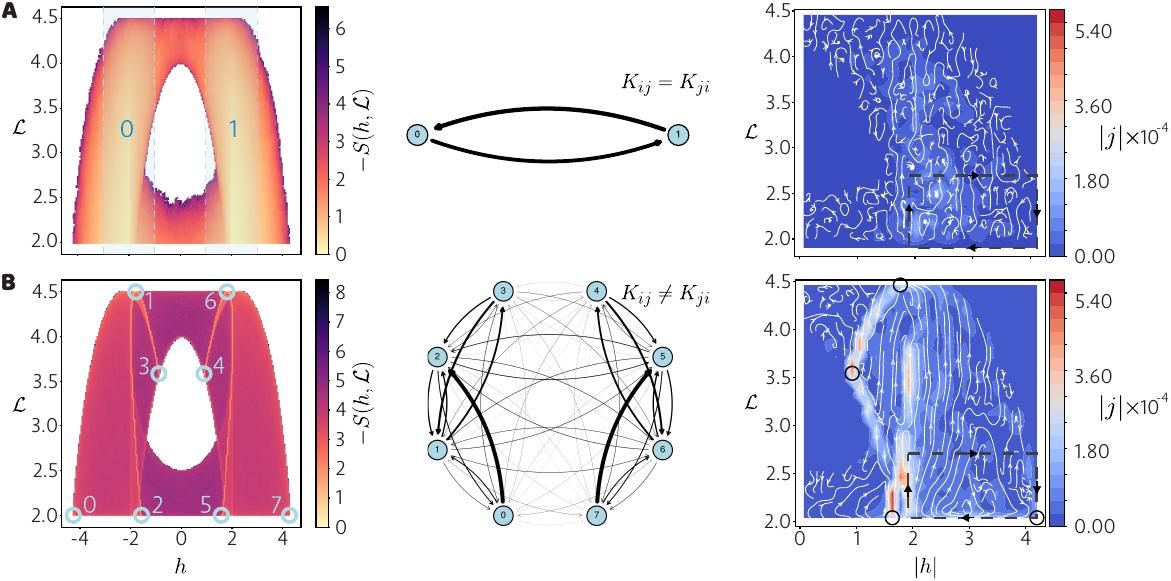}
    \caption{\textbf{Markov state model for entropic basin transition rates and breaking of detailed balance.} \textbf{A.} 2D entropic landscape, Markov state transition graph, and probability flux magnitude and streamlines map from near-equilibrium simulations $l_p=0.005$, $\epsilon=0.25$. Transition rates between states are symmetric near equilibrium, and probability fluxes are vanishing. \textbf{B.} Entropic landscape, Markov state transition graph, and probability flux plot for far-from equilibrium simulations $l_p=10$, $\epsilon=0.25$, showing highly asymmetric transition rates and substantial probability flux circulations. In both A and B, transition rates are represented on log scale by the widths of the arrows. Locations of the entropic landscape's local minima are labeled with larger circles than the threshold that was applied during graph computation for visibility.}
    \label{fig4:transition-graphs}
\end{figure}

Now we move beyond cage-breaking transitions and want to understand all transitions between the disk configurations that drive the entropic landscape's metastable basins. For this purpose it is insufficient to consider dynamics in the projected one-dimensional $h$--space, since many distinct geometric configurations of the disks can correspond to the same $h$ value. Therefore, we project to two dimensions by tracking $h$ as well as the length of the base of the triangle $\mathcal{L}$ for each timestep in our simulations (Fig. \ref{fig1:system-explanation}A for definition). This allows us to construct a histogram $n(h,\mathcal{L})$ and the 2D entropic landscape $-S(h,\mathcal{L})$ relative to the ground state in analagous fashion to Eq. (\ref{eq: entropic_landscape_def}), shown in the first column of Fig. \ref{fig4:transition-graphs}. 


The 2D landscape reflects similar trends as the 1D $h$ landscape, namely symmetry across the $h$ axis, wide basins of parameter space randomly explored at low persistence length, and sharply defined pathways at high persistence length (compare $l_p=0.005$ and $l_p=10$ landscapes in Fig. \ref{fig4:transition-graphs}A,B). There are large empty regions in $(h,\mathcal{L})$--space corresponding to ``forbidden" disk arrangements that require overlap. Local minima in the landscape (identified via local extrema detection) represent the most frequently sampled $(h,\mathcal{L})$ configurations of the system, comprising the metastable basins that are present in the 1D $h$ landscape along with some additional minima whose depths were previously insignificant when projected to 1D. The exact number of minima detected depends on the persistence length and the global threshold applied during extrema detection, see Supplementary Material.

Once the minima of the entropic landscape are identified, we use their locations as inputs to a new simulation with the same parameters to track transitions between the minima. This allows us to define a core-based Markov state model for the system, producing a transition rate matrix $K$ between the landscape's metastable minima (see Supplementary Fig. \ref{fig:SI-bootstrapped_Kij_pi}) \cite{Buchete2008CoarseDynamics,Prinz2011MarkovValidation, Nagel2019DynamicalModels}. This matrix is visualized in the state transition graphs of Fig. \ref{fig4:transition-graphs}, where the edge width from state $i$ to $j$ is proportional to the transition rate $K_{ij}$, normalized by the largest rate in the graph. 

The near-equilibrium graph has balanced transition rates between the two states. Far from equilibrium we find a fully-connected graph, consistent with the presence of noise in the system, for which there is a hierarchy of transition rates across several orders of magnitude. Additionally, several transitions are clearly asymmetric, i.e. $K_{ij}\neq K_{ji}$. The detailed balance condition is $p_i K_{ij} = p_j K_{ji}$ for any two states $i, j$, where $p_i$ is the probability to be in $i$. Violations of detailed balance imply a constant nonzero entropy production rate (EPR) for nonequilibrium steady-state systems, which can be computed for discrete-state systems using the Schnakenberg formula \cite{Schnakenberg1976NetworkSystems}
\be
\Phi = \dfrac{1}{2}\sum_{i,j}(p_i K_{ij} - p_j K_{ji})\log{\dfrac{p_i K_{ij}}{p_j K_{ji}}}.
\ee
The EPR of our system at $\ell_p=10$ is $\Phi = (3.7 \pm 0.1)\times 10^{-5} \; t_{\mathrm{red}}^{-1}$ (mean $\pm$ s.e.m. from 1000 bootstrap samples), which is significantly different from zero and verifies that our system breaks detailed balance at the level of the Markov state model. Meanwhile, near-equilibrium simulations with $\ell_p=0.005$ yield an EPR indistinguishable from zero after bootstrapping ($\Phi < 10^{-22} \; t_{\mathrm{red}}^{-1}$), consistent with detailed balance. See Supplementary Material for more details.

To complement our coarse-grained Markov model approach, we also demonstrate that detailed balance is broken at the finer-grained level of the full $(|h|,\mathcal{L})$--space when the persistence length is large by computing the probability flux field $\vec{j}(|h|, \mathcal{L})$ \cite{Battle2016BrokenSystems, Gnesotto2018BrokenReview, Cammann2021EmergentNavigation}, see Supplementary Material. The results are plotted in Fig. \ref{fig4:transition-graphs}. We observe sharp localization of the probability flux along the same entropically-favored pathways that are visible in the entropic landscapes, and significant flux loops are present for $\ell_p=10$, evidence of nonequilibrium steady-state behavior. To quantify the statistical robustness of the flux loops, we calculate the circulation of the flux field along the contours enclosed by the black boxes in Fig. \ref{fig4:transition-graphs}A and B, $\Omega = \oint \vec{j}\cdot d\vec{l}$. We find insignificant circulation for $\ell_p=0.005$ ($\Omega=(-3\pm 7)\times 10^{-5} \; t_{\mathrm{red}}^{-1}$, mean $\pm$ s.e.m. from 1000 bootstrap samples) and significant circulation for $\ell_p=10$ ($\Omega=(5.4\pm 0.5)\times 10^{-4} \; t_{\mathrm{red}}^{-1}$), confirming the breaking of detailed balance in $(|h|,\mathcal{L})$--space far from equilibrium.


\section*{Conclusion}
In summary, we studied a minimal model of caging dynamics consisting of three active disks under circular confinement. Although simple, this system captures the interplay between self-propulsion and steric repulsion that underlies clustering and motility-induced aggregation in bulk active matter \cite{Tailleur2008StatisticalBacteria, Fily2012AthermalAlignment, Cates2012DiffusivePhysics, Levis2014ClusteringDisks, Cates2015Motility-inducedSeparation, Nie2020StabilityParticles, Keta2022DisorderedParticles}. In our confined geometry, these ingredients give rise to transient clustered structures on the boundary, which persist for times of order $\mathcal{O}(\tau_p)$ and have a profound impact on the system's rearrangement dynamics. By interpreting these structures as metastable minima of an entropic landscape in $h$--space \cite{Hunter2012Free-energyDisks, Du2016EnergyModel, Weeks2020VisualizingDisks}, we characterize the transitions between metastable configurations, including the cage-breaking events in which particles exchange positions. Persistent motion is found to deform the bistable equilibrium landscape into a multistable one, as well as driving nonequilibrium probability currents and breaking detailed balance both in the continuous dynamics and the coarse-grained transition graph. 

These results establish a link between active caging dynamics and landscape-based concepts familiar from glass physics, such as shear transformation zones and inherent structures \cite{Spaepen1977MicroscopicFlow, Patinet2016ConnectingSolids, Stillinger1995AFormation, Speedy1994TwoBox, Bowles1999FiveBox}. There, landscape pictures have proven powerful but remain difficult to ground microscopically; our minimal model suggests that explicitly geometric, few-body approaches offer a complementary route to making such pictures more precise. Conversely, these concepts provide useful interpretive language for active caging in bulk systems where similar landscape thinking is only beginning to be applied \cite{Chaki2020EscapeDiffusion, Klongvessa2019ActiveSelf-Propulsion}. While important bulk glass concepts such as dynamic heterogeneity and fragility are absent or ill-defined in such a few-body setting, the model nevertheless exhibits elementary ingredients of active glassy relaxation in a form that can be analyzed quantitatively, including optimal cage breaking at an intermediate persistence length set by the system's geometry \cite{Debets2021CageMatter}.

Collectively, our work shows that low-dimensional entropic landscapes can provide a useful bridge between nonequilibrium active matter and the language of metastability, basin escape, irreversibility, and geometry-driven optimization, while also offering a minimal setting in which emergent memory effects may be identified and quantified.



\section*{Acknowledgements}
The authors thank Peter Sollich, Eric R. Weeks, Ludovic Berthier, and Corentin Coulais for fruitful discussions. M.J. acknowledges the ERC grant no.~``2023-StG-101117025, FluMAB", NWO (Dutch Research Council) under the MIST program and the Vidi project Living Levers with No. 279 21239, financed by the Dutch Research Council (NWO).

\section*{Competing interests}
The authors declare no competing interests.

\section*{Author contributions}
All authors conceived of the project. J.P. performed the simulations and prepared the first draft of the manuscript. J.P. and N.S. analyzed the data. J.P., N.S., and M.J. designed the figures. M.J. acquired funding. All of the authors contributed to the review of the manuscript before submission for publication and approved the final version.



\printbibliography


\pagebreak

\pagebreak
\section*{Supplementary Material}
\subsection*{A. Simulation Details}
\subsubsection*{Langevin dynamics}
We initialize simulations with three distinguishable quasihard disks of effective radius $r_{\mathrm{eff}}$ placed with random non-overlapping positions within a fixed quasihard-walled circular confinement of radius $R=3\,r_{\mathrm{eff}}+\epsilon$. Each particle $i$ is described by the following athermal overdamped Langevin equation \cite{Debets2021CageMatter, Fily2012AthermalAlignment}: 
\be
    \dot{\mathbf{r}}_{i}=\zeta^{-1}(\mathbf{F}_i + \mathbf{f}_i), \label{eq: SI-eq-of-motion}
\ee
where $\mathbf{r}_i$ denotes the position of particle $i$, $\zeta$ is the damping coefficient, and $\mathbf{F}_i$ and $\mathbf{f}_i$ are the interaction and self-propulsion forces acting on particle $i$, respectively.  The damping coefficient is set to unity in our simulations $\zeta=1$. The interaction force $\mathbf{F}_i=-\sum_{j\neq i, \mathrm{wall}}\nabla_i V(r_{ij})$ sums over all particle/particle and particle/wall pairs and is derived from a WCA potential,
\be
    V(r_{ij})=
\begin{cases}
4E_{\mathrm{int}}\left[\left(\dfrac{\sigma}{r_{ij}}\right)^{12} - \left(\dfrac{\sigma}{r_{ij}}\right)^{6} \right] + E_{\mathrm{int}} & \text{for } r_{ij}\leq 2^{1/6}\sigma, \\
0 & \text{for } r_{ij} > 2^{1/6}\sigma,
\end{cases}
\ee
where $\sigma = 2\cdot 2^{-1/6}$ is the length parameter of the potential, and we generally choose a large interaction energy of $E_{\mathrm{int}}=40$ to make the disks behave as approximately hard disks. For the optimum caging time results we used $E_{\mathrm{int}}=1000$, as we found that the location of the $\ell_p^{\mathrm{min}}$ optimum depends strongly on the value of $E_{\mathrm{int}}$, beginning at $\ell_p^{\mathrm{min}}\approx 0.89$ for $E_{\mathrm{int}}=40$ and increasing towards $r_{\mathrm{eff}}=1$ as $E_{\mathrm{int}}$ becomes far larger than any other energy scale in the system. The WCA potential has been shown in previous studies to provide a good approximation for the interactions of colloidal hard disks, with an effective hard-disk diameter of the truncation distance $\sigma_{\mathrm{eff}}=2^{1/6}\sigma$ at low temperature \cite{Andersen1971RelationshipForces, Heyes2006EquationFluid, Kawasaki2010FormationLiquid, Filion2011SimulationPersists, Ni2015TunableSpheres, Attia2022ComparingLine}. In defining the WCA potential we make the nonstandard choice $\sigma = 2\cdot 2^{-1/6}$ as opposed to $\sigma=2$ for two reasons: such that the effective hard-disk radius of the particles is $r_{\text{eff}}=1$, and to facilitate comparison with previous literature on the equilibrium version of this model \cite{Hunter2012Free-energyDisks, Du2016EnergyModel, Weeks2020VisualizingDisks}. Interactions with the walls are handled via the same WCA potential as $V_{\mathrm{wall}}(i,j)=V_{\mathrm{WCA}}(r_{ij})$, in which $r_{ij}$ is the minimal distance between the center of the particle $i$ and points on the circular confinement $j$ plus one effective radius $r_{\mathrm{eff}}$, so that the confinement boundary represents the ``edge" of the wall potential.

Since Eq. (\ref{eq: SI-eq-of-motion}) is athermal, the only source of noise in the particle dynamics is due to the self-propulsion force. For the self-propulsion force, we treat the disks as active Brownian particles (ABPs) with a constant magnitude self-propulsion force $\mathbf{f}_i = f\mathbf{e}_i$ whose orientation $\mathbf{e}_i$ undergoes a rotational diffusion process
\be
    \dot{\mathbf{e}}_i = \mathbf{\chi}_i \times \mathbf{e}_i,
\ee
subject to a Gaussian noise process with zero mean and variance $\langle\mathbf{\chi}_i(t)\mathbf{\chi}_j(t')\rangle_{\mathrm{noise}}=2D_{r}\mathbf{I}\delta_{ij}\delta(t-t'),$ whose amplitude is determined by the rotational diffusion coefficient $D_r$. Neglecting particle interactions, ABP dynamics lead to a persistent random walk characterized by a persistence time $\tau_p = (2D_r)^{-1}$ and a persistence length of $\ell_p=\zeta^{-1}f\tau_p$, which serve as natural control parameters for the level of activity throughout our study.

Although ABPs are not generically ergodic \cite{Basu2025ErgodicityPotential} — a bounded propulsive force cannot overcome arbitrarily deep confining potential wells — our model uses a purely repulsive quasi-hard WCA potential with no local minima or attractive basins, and we fix $\epsilon > 0$ such that the confinement geometry permits cage-breaking transitions. Therefore the full configuration space is dynamically accessible, and the system reaches a nonequilibrium steady state in which ensemble averages equal time averages. Thus we are justified in using a single long trajectory to estimate $n(h)$. 

To simulate the model's dynamics over the time interval $[0, T_{\mathrm{max}}]$, we cast Eq. (\ref{eq: eq-of-motion}) as an It\^{o} stochastic differential equation and apply a strong order 1 Runge-Kutta integration method, which produces a discretized trajectory over the interval in a specified number of time steps $n_{\mathrm{samp}}$ \cite{Roberts2012ModifyEquations}. For a given interaction energy $E_{\mathrm{int}}$, we always choose $n_{\mathrm{samp}}$ and $T_{\mathrm{max}}$ such that the timestep $dt=T_{\mathrm{max}}/n_{\mathrm{samp}}$ is fixed, with $dt=10^{-4}$ for simulations with $E_{\mathrm{int}}=40$ and $dt=4\cdot 10^{-6}$ for $E_{\mathrm{int}}=1000$. This is far smaller than other timescales in the system and prevents substantial overlaps between disks or the confinement in adjacent time steps. All results are presented in reduced units where $r_{\mathrm{eff}}, E_{\mathrm{int}},$ and $t_{\mathrm{red}}\equiv\zeta r_{\mathrm{eff}}^2/E_{\mathrm{int}}$ represent the units of length, energy, and time, respectively. All simulation code is available on Github \cite{simulations_github}.

\subsubsection*{Sampling stable configurations at infinite persistence}
In the $\tau_p \to \infty$ limit, a single disk with a randomly-initialized active force vector directed at an angle $\theta_1$ will move in a straight line in that direction until it contacts the boundary, where it crawls along the wall until its force direction points radially outward at wall position $\psi=\theta_1$, i.e. parallel to the outward normal vector at the point of contact. In this position the disk's active force is balanced by the contact force from the wall. 

For multiple disks, the final position can be arrested before disk 1 reaches $\psi=\theta_1$ if it contacts another disk. In this case the final position is determined by a force balance between the active and contact forces between the disks and walls. In the case of two disks in contact, this force balance occurs when the angles of the active forces relative to the lines from the confining center to the disk centers are equal and opposite, $\alpha = \beta$. Motivated by this simple picture, we posit that simulations at very large but finite $\tau_p$ will follow essentially the same dynamics, crawling along the walls until they reach a force-balanced static configuration, with rare jumps to other static configurations due to the finite rotational diffusion of their activity vectors. Hence, to approximate the entropic landscape as the persistence time approaches to infinity, we need to sample the $h$ values over many randomly generated force-balanced stable configurations of the disks.

\begin{figure}[!tbph]
    \centering
    \includegraphics[width=0.7\textwidth]{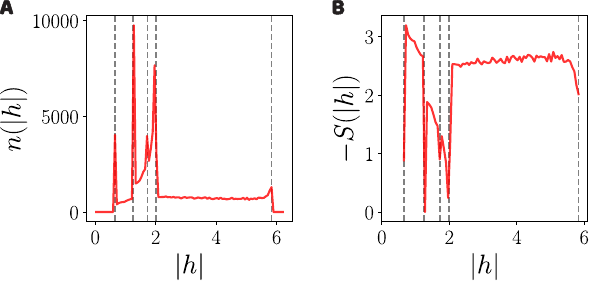}
    \caption{\textbf{A.} The distribution $n(|h|)$ for boundary-initialized stable configuration simulations, $\epsilon=1$. Theoretical peak locations at $h_{1}, \ldots, h_5$ are marked with vertical dashed lines. \textbf{B.} The corresponding entropic landscape, where the peaks in the distribution become local entropic minima. Note the lack of an entropic barrier at $h=0$, as by only sampling stable configurations we do not have any cage breaking transitions. Here we plot the values at the bin center locations of the $n(h)$ histograms, and slight apparent discrepancies between the simulated entropic minima locations and the theoretical predictions are due to finite-width bins.}
    \label{fig:SI-stable-configs}
\end{figure}

We sample the $h$ distribution of disks at infinite persistence with the following procedure: we initialize the disks at random non-overlapping positions on the boundary (disk center at distance $R-r_{\mathrm{eff}}$ from the confinement center) with fixed random activity directions drawn from a uniform distribution
\begin{equation}
    P(\theta_n)=
\begin{cases}
\dfrac{1}{2\pi} & \text{for } 0\leq \theta_n < 2\pi \\
0 & \text{for } \theta_n < 0 \text{ or } \theta_n > 2\pi,
\end{cases}
\quad n \in [1,2,3],
\end{equation}

run the same Langevin dynamics as in finite-persistence simulations until the disks reach a stable configuration ($h$ stayed within the range $\pm0.0005$ for 1000 timesteps $dt$), and record the $h$ value of that configuration. This process is repeated over 100 random non-overlapping initial arrangements of the disks on the confinement boundary, and for each arrangement 1000 random sets of activity directions for the three disks were ran, for a total of one hundred thousand total configurations sampled to estimate the histogram $n(h)$. From there the infinite-persistence entropic landscape is computed as in all other simulations, using Eq. \ref{eq: entropic_landscape_def} from the main text. See Fig. \ref{fig:SI-stable-configs}.

The choice of random boundary initialization was physically motivated by the disk dynamics in large finite $\tau_p$ simulations, where disks spend large amounts of time in stable configurations on the boundary and displace themselves at random intervals dependent on $\tau_p$. Empirically we found almost no difference in the $h$-distribution when initializing disks on the boundary or throughout the entire confinement area.

When searching through the disk configurations that produce the peaks of the $n(h)$ distributions in Fig. \ref{fig:SI-stable-configs}, we found that disks were consistently arranged as shown at the top of Fig. \ref{fig2:entropic-landscape}A. It is a simple geometry problem to compute the $h$ values for hard disks of radius $r_{\mathrm{eff}}$ in these particular arrangements within a circular confinement of radius $R$, with the results as follows: 

\begin{align}
    h_1 &= (R-r_{\mathrm{eff}})\cdot \left[1-\cos{\left(2\arcsin\left(\dfrac{r_{\mathrm{eff}}}{R-r_{\mathrm{eff}}}\right)\right)}\right] \notag \\
    h_2 &= 2r_{\mathrm{eff}}\cdot \sin{\left(\pi-2\arccos{\left(\dfrac{r_{\mathrm{eff}}}{R-r_{\mathrm{eff}}}\right)}\right)} \notag \\
    h_3 &= \sqrt{3}r_{\mathrm{eff}} \notag \\
    h_4 &= 2r_{\mathrm{eff}} \notag \\
    h_5 &= R-r_{\mathrm{eff}} + \sqrt{(R-r_{\mathrm{eff}})^{2} - r_{\mathrm{eff}}^2} \label{eq: minima_locations}
\end{align}
These minima locations are numbered in increasing values of $h$. Note that since $R=3r_{\mathrm{eff}}+\epsilon$, Eqs. (\ref{eq: minima_locations}) give immediate predictions for the locations of each entropic landscape minimum as a function of $\epsilon$, $h_{i}(\epsilon)$. These minima are plotted as vertical dashed lines in the $\epsilon=1$ entropic landscape plot of the main text Fig. \ref{fig2:entropic-landscape}A and in Fig. \ref{fig:SI-stable-configs} for the $\epsilon=1$ stable configurations distributions, and their dependence on confinement $h_{i}(\epsilon)$ is plotted as the solid curves in Fig. \ref{fig2:entropic-landscape}B.

Our procedure is not expected to perfectly reproduce the entropic landscape explored by simulations at large but finite $\tau_p$, since for those simulations we record the $h$ values at every time step. Indeed, this procedure of recording stable configurations amounts to mapping out the system's landscape of stable attractors, as opposed to the geometry of trajectories between these attractors. As a consequence we lose dynamic information such as statistics at $h=0$ of cage breaking transitions with such an approach (see Fig. \ref{fig:SI-stable-configs}B). However, the quite accurate reproduction of all other features of the entropic landscape as the finite $\tau_p$ simulations that we observed in Fig. \ref{fig2:entropic-landscape}, including the locations of the local minima and a similar Wasserstein-1 distance $W_{\mathrm{stable}}$ as the fit $W_{1}^{\infty}$ to our finite $\tau_p$ data, gives us confidence that the minima in the entropic landscape are driven by stable configurations of disks in contact on the boundary.

\subsubsection*{Wasserstein-1 distance calculation - analytic $h$ distribution at equilibrium}
For hard disks at equilibrium, Hunter and Weeks derived the analytic function $n_{\mathrm{eq}}(h)$ and therefore the thermodynamic free energy $F(h)=-\log(n_{\mathrm{eq}}(h)/n_0)$ in the Appendix of \cite{Hunter2012Free-energyDisks}. In short, one can compute $n_{\mathrm{eq}}(h)$ by integrating over the space of allowed configurations $\Omega$ of the three disks while maintaining a fixed $h$. In general, this can be written

$$n_{\mathrm{eq}}(h)=\int_{\Omega}d\vec{r}_1 d\vec{r}_2 d\vec{r}_3 \delta[h-H(\vec{r}_1,\vec{r}_2,\vec{r}_3)],$$

where the function $H(\vec{r}_1,\vec{r}_2,\vec{r}_3)$ calculates the value of $h$ given the coordinates of the three disks and $\delta(h)$ is the Dirac delta function. The expression above can be reduced to a 1D integral with four geometrically distinct cases to consider given the relative locations of the disks, each of which we then integrate numerically. This distribution serves as the ``ground truth" with respect to which we compute the Wasserstein-1 distance of our simulated entropic landscape. In other words, we accumulate the histograms $n_{\mathrm{sim}}(h)$ over our simulations at each persistence time $\tau_p$, then compute the Wasserstein-1 distance from $n_{\mathrm{eq}}(h)$ to $n_{\mathrm{sim}}(h)$.




\subsection*{B. Entropic landscape and Wasserstein-1 distance at $\epsilon=0.25$}

\begin{figure}[!tbph]
    \centering
    \includegraphics[width=0.7\textwidth]{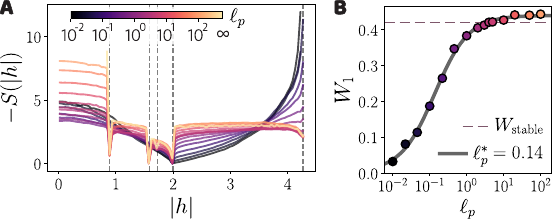}
    \caption{\textbf{A.} The entropic landscape as a function of persistence length $\ell_p$ for $\epsilon=0.25$, with the static configurations landscape plotted as the curve at $\ell_p=\infty$. \textbf{B.} The corresponding Wasserstein-1 distance of the simulated distributions to the analytic distribution at equilibrium (Ref. \cite{Hunter2012Free-energyDisks}).}
    \label{fig:SI-landscape-epsilon=0.25}
\end{figure}

As a point of comparison with the $\epsilon=1$ entropic landscapes plot of Fig. \ref{fig2:entropic-landscape} in the main text, we provide here in Fig. \ref{fig:SI-landscape-epsilon=0.25} the entropic landscape and Wasserstein-1 distance for a tighter confinement of $\epsilon=0.25$. Both plots show the same trends as in the main text, namely the same five entropic landscape minima, a nonmonotonic dependency of the entropic barrier height on the persistence time, and a Wasserstein distance that increases monotonically with persistence. However, in Fig. \ref{fig:SI-landscape-epsilon=0.25}B the inflection point $\ell_p^{*}=0.14$ of the sigmoid function fit to the data and the Wasserstein-1 distances $W_{\mathrm{stable}}=0.42$ (infinite-persistence stable configurations distribution) and $W_1^{\infty}=0.44$ (sigmoid function fit) are smaller than the corresponding quantities for $\epsilon=1$ in the main text, for which we found $\ell_p^{*}=0.36$, $W_{\mathrm{stable}}=0.55$, and $W_1^{\infty}=0.59$, respectively. This indicates that the particular activity level at which the landscape deviates substantially from near-equilibrium behavior is a function of the level of system confinement, rather than being a universal number. It may be interesting to systematically explore this transition to far-from-equilibrium behavior across various system densities in a future study.

\subsection*{C. Caging time dependence on confinement flattens with increasing persistence length}
\begin{figure}[!tbph]
    \centering
    \includegraphics[width=0.5\textwidth]{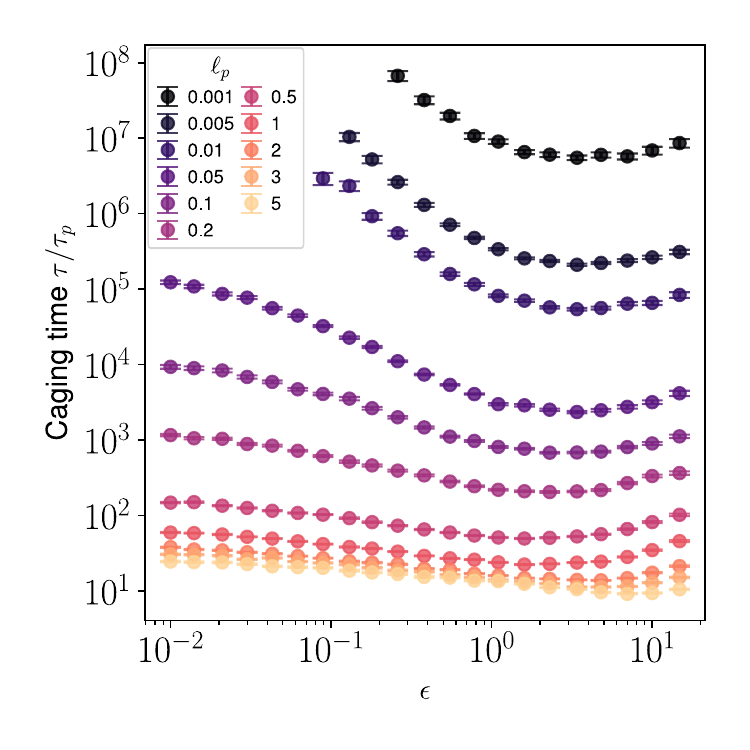}
    \caption{Cage time becomes nearly constant with respect to the confinement parameter $\epsilon$ for large $\ell_p$ and small $\epsilon$, after normalization by the persistence time $\tau_p$. In these limits the geometry of the cage is effectively fixed, and the disks simply need to reorient some average number of times $N_{\mathrm{attempts}}$ before cage breaking.}
    \label{fig:SI-normalized-cage-times}
\end{figure}

As the persistence length of our confined active disks increases, they gradually transition from diffusive exploration to ballistic exploration limited by geometry. In particular, highly persistent disks travel in nearly straight-line paths with a constant heading, only deviating during contact with confinement or other disks. If in this motion they failed to produce a cage breaking event, they have to wait time $\mathcal{O}(\tau_p)$ to reorient and try again. This repeats until there is a successful crossing. If the geometry at all permits a crossing (which it does for all $\epsilon > 0)$, then the disk essentially has to explore angles until it happens upon just the right solid angle to pass between the other two disks. Thus, in the large $\ell_p$ limit at small $\epsilon$ we should observe a linear scaling of the caging time $\tau$ with $\tau_p$,
\be
    \tau \sim N_{\mathrm{attempts}}\cdot \tau_p,
\ee
where the typical number of attempts before a zero-crossing $N_{\mathrm{attempts}}$ is a purely geometric quantity, i.e. roughly constant for $\epsilon\to 0$. If this argument is right, then a plot of $\tau/\tau_p$ should show no dependence on $\epsilon$ in the caging regime $\epsilon \lesssim 1$ at high persistence length $\ell_p$, instead collapsing to a straight line. As demonstrated in Fig. \ref{fig:SI-normalized-cage-times}, the rescaled caging time indeed becomes substantially less sensitive to $\epsilon$ as $\ell_p$ increases.

\subsection*{D. Quantifying broken detailed balance}
In the main text, we found that particle activity drives the entropic landscape of our confined three-particle system from bistable to multistable, with the additional states arising from configurations of the particles clustered on the boundary. These states represent the minima in the 2D (negative) entropic landscape of Fig. \ref{fig4:transition-graphs}B. To test for activity-driven violation of detailed balance in the dynamics, we studied two levels of coarse-graining of our system: a core-based Markov state model focused solely on transitional dynamics between the metastable states, and the system dynamics in the full $(|h|, \mathcal{L})$-space. 

\subsubsection*{Markov state model and state transition graph}
Our core-based Markov state model is inspired by work on transitions between metastable configurations in biomolecular dynamics \cite{Buchete2008CoarseDynamics, Prinz2011MarkovValidation, Nagel2019DynamicalModels}, and consists of the residence times and transition rates between the local minima in the entropic landscape. For near-equilibrium simulations there are only two broad minima at $h=\pm 2$. Further from equilibrium, we extract the minima locations by running a long $\mathcal{O}(10^{10})$ timestep simulation to obtain an accurate $n(h,\mathcal{L})$ histogram. Dividing each bin $\alpha$ in this histogram by the total number of counts gives us the probability mass function
\be
p_{\alpha}(h,\mathcal{L})=\dfrac{n_{\alpha}(h,\mathcal{L})}{\sum_{\alpha}n_{\alpha}(h,\mathcal{L})}.
\ee

\begin{figure}[!tbph]
    \centering
    \includegraphics[width=0.7\textwidth]{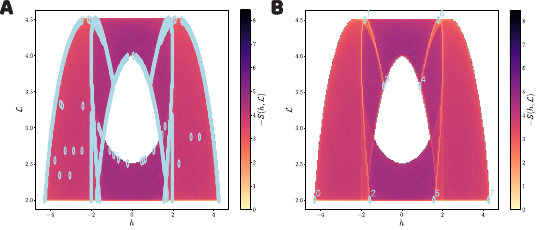}
    \caption{\textbf{A.} The negative entropic landscape $-S(h,\mathcal{L})$ for $\ell_p=10$. On top of the landscape in light blue markers is plotted the candidates for the local extrema resulting from a 1D peak detection in the conditional distributions $p_{\alpha}(h|\mathcal{L}=\mathcal{L}_{j})$ for each $\mathcal{L}_j$. The constraint curves generated by the disks arranged on the boundary are identified by our procedure as continuous ridges of local maxima in the probability distribution (minima in the negative entropic landscape). \textbf{B.} The remaining peak locations after filtering the candidates to only preserve those that fall above a global threshold of $0.08 \cdot \max_{h \mathcal{L}}\,p_{\alpha}(h,\mathcal{L})$. Note the symmetry of the peak locations with respect to the sign of $h$.}
    \label{fig:SI-peakfinding}
\end{figure}

The nonequilibrium system generically has probability concentrated along sharply-defined ridges in $(h, \mathcal{L})$-space (the constraint curves generated by disks arranged on the boundary) rather than in isolated point-like peaks. A naive 2D local maximum filter with a small window would therefore find many spurious maxima along a flat ridge, whereas we are interested only in the dominant local maxima corresponding to distinct metastable states for our Markov model. Therefore we run a two-stage local extrema identification on $p_{\alpha}(h,\mathcal{L})$, consisting first of 1D peak detection in the conditional distributions $p_{\alpha}(h|\mathcal{L}=\mathcal{L}_{j})$ for each value of $\mathcal{L}_{j}$ (Fig. \ref{fig:SI-peakfinding}A). Each peak must meet a relative prominence threshold within its slice of $\geq 0.01\max{p_{\alpha}(h|\mathcal{L}=\mathcal{L}_{j})}$ to be considered as a local maximum candidate. This stage primarily identifies the ridges in $(h, \mathcal{L})$-space where probability is concentrated, as well as a few spurious local peaks due to the finite simulation runtime. In the second stage, we discard any candidate at $(h_i, \mathcal{L}_j)$ whose probability $p_{\alpha}(h_i, \mathcal{L}_j)$ falls below a global threshold of $\theta \cdot \max_{h \mathcal{L}}\,p_{\alpha}(h,\mathcal{L})$, where $\theta = 0.08$ was chosen to retain only the dominant maxima on each ridge, collapsing the many conditional-slice candidates to a small number of physically meaningful basins (Fig. \ref{fig:SI-peakfinding}B).

Identified basin locations for a given $\ell_p$ serve as inputs to a second $\mathcal{O}(10^{10})$ timestep simulation, where transitions into and out of the minima are tracked to construct the Markov state model. We count a transition as having occurred if the $(h, \mathcal{L})$ value of the system passes within a small arbitrarily-defined distance threshold of $r=0.05$ from the minima locations, also recording the simulation time elapsed since the previous transition. For all other $(h, \mathcal{L})$ values the system is considered ``in transit" between these minima. For the near-equilibrium simulation with $\ell_p=0.005$, the broadness of the entropic landscape in the $\mathcal{L}$-coordinate means that we only track transitions in the $h$-coordinate. Specifically, we define target regions $|h| \in (1,3)$ centered on each minimum at $h=\pm 2$, and record a transition when the system crosses from one region to the other (see Fig. \ref{fig4:transition-graphs}A for a plot of these regions compared with the entropic landscape).

The result of such a simulation is a matrix $A(t_{n})$ of size $N\times 3$, where $N$ is the total number of transitions recorded during the simulation. $A(t_{n})$ stores in each row the previous state before the transition, the state that was transitioned into, and the time interval over which the transition took place,

\be
A(t_{n})=
\begin{pmatrix}
s_1 & s_2 & t_{1,2}\\
s_2 & s_3 & t_{2,3}\\
& \ldots & \\
s_{N-1} & s_{N} & t_{N-1, N} 
\end{pmatrix}.
\ee

By counting all rows of $A(t_n)$ containing a transition from $i$ to $j$ we obtain the number of transitions $N_{i\to j}$, and summing the transition time intervals in these rows gives us the residence times $T_i$. The maximum likelihood estimator (MLE) of the probability of residing in state $i$ is
\be\label{eq: SI-MLE-prob_state_i}
p_i = \dfrac{T_i}{\sum_i T_{i}}
\ee
while the MLE for the transition rate from state $i$ to state $j$ is 
\be\label{eq: SI-MLE-transition-rate-Kij}
K_{ij}=\dfrac{N_{i\to j}}{T_i}.
\ee
We define the diagonal entries as $K_{ii}=-\sum_{j \neq i}K_{ij}$ such that each row sums to zero, as is standard for continuous-time Markov chains \cite{Nartallo-Kaluarachchi2025Coarse-grainingChains}.

To estimate uncertainties on $p_i$ and $K_{ij}$, we follow the bootstrap sampling procedure introduced by Battle et al. \cite{Battle2016BrokenSystems}, reviewed in more depth in \cite{Gnesotto2018BrokenReview}. For each of 1000 bootstrap replicates we resample $A(t_n)$ by drawing $\lfloor N/m \rfloor$ sets of $m$ consecutive rows with replacement, producing a resampled trajectory of length $\sim N$. We chose $m=2$ following Battle et al. \cite{Battle2016BrokenSystems} to account for possible pairwise correlations, but we tested $m=1, 2, 5,$ and $10$ and our results did not depend on this choice, implying that our metastable state-based coarse-grained system is Markovian. For each bootstrapped trajectory we calculate $p_i$ and $K_{ij}$ using Eqs. (\ref{eq: SI-MLE-prob_state_i}) and (\ref{eq: SI-MLE-transition-rate-Kij}), which allows us to compute the mean and standard deviation of our bootstrapped estimates.

\begin{figure}[!tbph]
    \centering
    \includegraphics[width=0.7\textwidth]{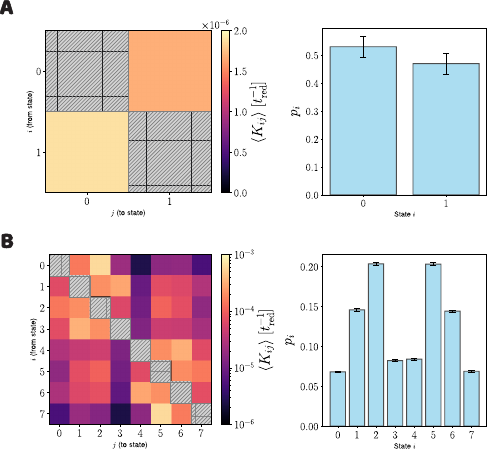}
    \caption{\textbf{A.} Transition rates and state occupation probabilities for $\ell_p=0.005$.  \textbf{B.} Transition rates and state occupation probabilities for $\ell_p=10$. In both transition rate plots, the diagonal entries $K_{ii} = -\sum_{j\neq i} K_{ij}$ give the total escape rate from state $i$ and are negative by construction; we hatch them as they lie outside the colormap range.}
    \label{fig:SI-bootstrapped_Kij_pi}
\end{figure}

We depict the mean transition rates $\langle K_{ij}\rangle$ and occupation probabilities $p_i$ with bootstrapped error bars (standard deviation) for our near-equilibrium $\ell_p=0.005$ and far-from-equilibrium $\ell_p=10$ simulations in Fig. \ref{fig:SI-bootstrapped_Kij_pi}. Since the diagonal entries $K_{ii}$ are negative by construction, we hatch them since they are not transition rates between distinct states and fall outside the colormap range. The same transition rates are visualized in Fig. \ref{fig4:transition-graphs} of the main text in a state transition graph, whose edge width from state $i$ to $j$ is proportional to the transition rate $K_{ij}$, normalized by the largest rate in the graph. 

\begin{figure}[!tbph]
    \centering
    \includegraphics[width=0.7\textwidth]{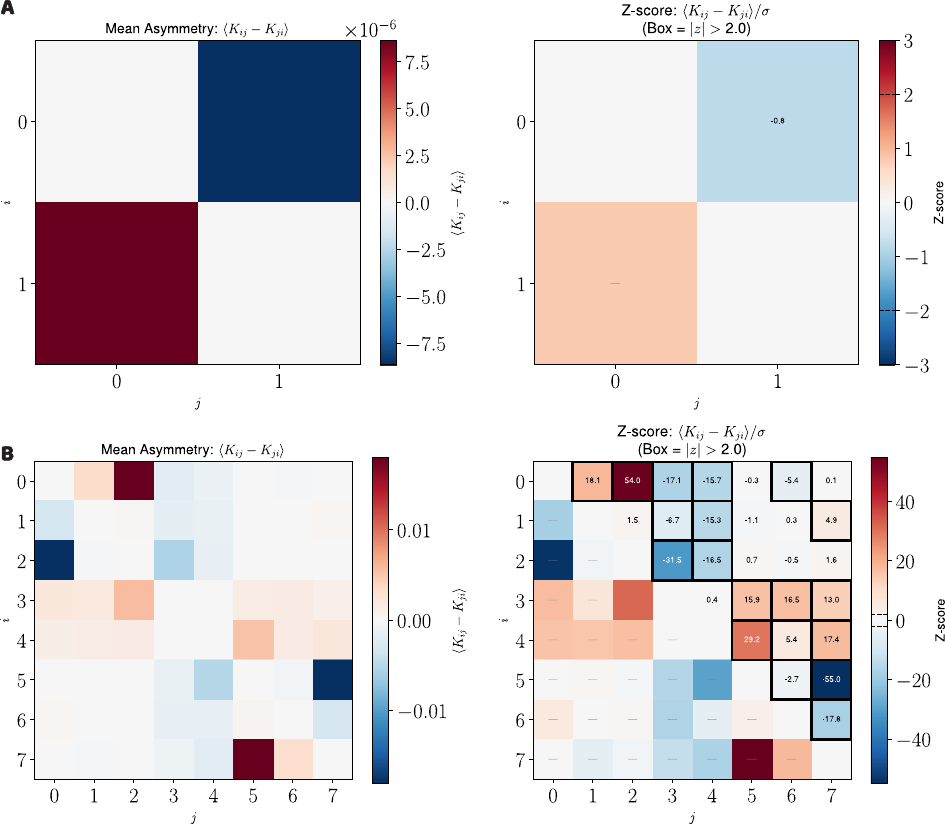}
    \caption{\textbf{A.} Mean asymmetry and $z$-score for the near-equilibrium simulations at $l_p=0.005$. \textbf{B.} Mean asymmetry and $z$-score for the near-equilibrium simulations at $l_p=10$. Matrix elements having $z$-scores of $|z|>2$ are highlighted in thick boxes.}
    \label{fig:SI-bootstrap-asymmetry}
\end{figure}

From our transition matrices we compute the mean matrix asymmetry $\langle K_{ij}-K_{ji}\rangle$, averaged over all pairs of states $i,j$ (Fig. \ref{fig:SI-bootstrap-asymmetry}). If the system is at equilibrium, state transitions are balanced and the matrix asymmetry equals zero. To be sure that an observed nonzero matrix asymmetry is not a statistical effect, we calculate the $z$-score on the matrix asymmetry $\langle K_{ij}-K_{ji}\rangle/\sigma$, quantifying confidence in rejecting the null hypothesis that the asymmetry is zero. We find for the highly persistent graphs enormous $z$-scores of up to 75. By contrast, the $l_p=0.005$ near-equilibrium simulations have a $z$-score of 0.8, which is not statistically different than zero and is consistent with balanced transition rates.

\begin{figure}[!tbph]
    \centering
    \includegraphics[width=0.4\textwidth]{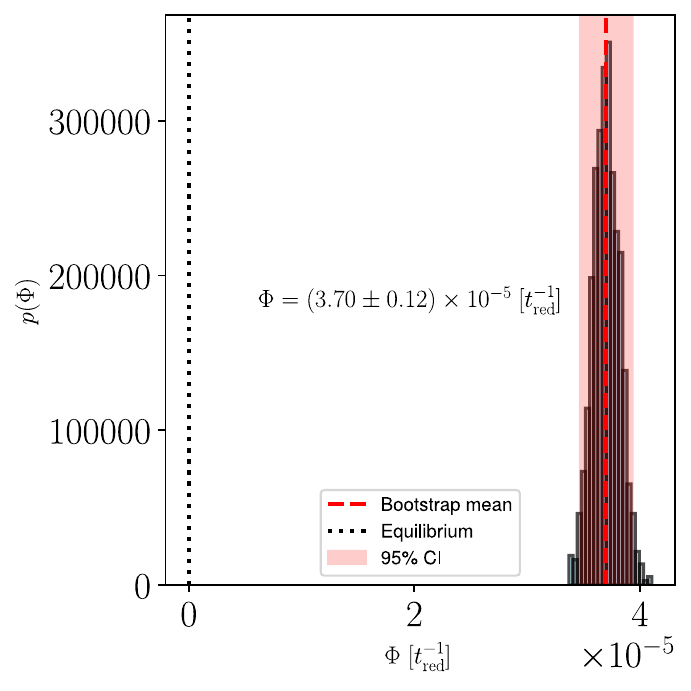}
    \caption{The entropy production rate $\Phi$ for our Markov state model at $\ell_p=10$ is significantly different from zero, demonstrating the breaking of detailed balance for dynamics on the state transition graph. The distribution of $\Phi$ is estimated over 1000 bootstrap realizations, with the 95\% confidence interval shown as a shaded red box. The error value of $0.12 \times 10^{-5} \; [t_{\mathrm{red}}^{-1}]$ listed in the figure is the bootstrapped standard error of the mean.}
    \label{fig:SI-entropy-production-rate}
\end{figure}

For state transitions on a network, the detailed balance condition is $p_i K_{ij} = p_j K_{ji}$ for any two states $i, j$ \cite{Nartallo-Kaluarachchi2025Coarse-grainingChains}. Systems that break this condition have a nonzero entropy production rate (EPR), which can be computed using the Schnakenberg formula \cite{Schnakenberg1976NetworkSystems}
\be
\Phi = \dfrac{1}{2}\sum_{i,j}(p_i K_{ij} - p_j K_{ji})\log{\dfrac{p_i K_{ij}}{p_j K_{ji}}}.
\ee
Using our bootstrapped values of $p_i$ and $K_{ij}$, we compute the EPR for our $\ell_p=10$ simulations, obtaining the distribution $p(\Phi)$ of estimated EPR values shown in Fig. \ref{fig:SI-entropy-production-rate}. The result of $\Phi = (3.7 \pm 0.1)\times 10^{-5} \; t_{\mathrm{red}}^{-1}$ (mean $\pm$ s.e.m. from 1000 bootstrap samples) is significantly different from zero, confirming the breaking of detailed balance on the level of our coarse-grained Markov model. Meanwhile, we found in Figs. \ref{fig:SI-bootstrapped_Kij_pi}A and \ref{fig:SI-bootstrap-asymmetry}A that occupation probabilities and transition rates are balanced in our $\ell_p=0.005$ simulations, and hence its EPR should be zero. After bootstrapping we indeed find $\Phi = (1.9 \pm 2.4)\times 10^{-23} \; t_{\mathrm{red}}^{-1}$ (mean $\pm$ s.e.m. from 1000 bootstrap samples), consistent with zero EPR and detailed balance.

\begin{figure}[!tbph]
    \centering
    \includegraphics[width=0.85\textwidth]{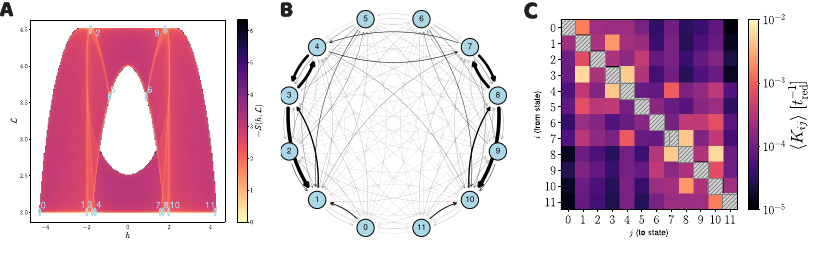}
    \caption{\textbf{A.} 2D entropic landscape with basin assignments for $\ell_p=\sigma/2$. \textbf{B.} Corresponding Markov state transition graph. \textbf{C. } Transition rates for the Markov model in B. Note the relatively uniform coloration compared to the $\ell_p=10$ transition rates in Fig. \ref{fig:SI-bootstrapped_Kij_pi}B.}
    \label{fig:SI-optimal-taup-network}
\end{figure}

The number of minima identified by our two-stage local extrema identification depends on the fitting procedure through the choice of thresholds in each stage. As discussed above, we found empirically that a relative prominence threshold of 0.01 and global probability threshold of $\theta=0.08$ worked well for our system to identify the most dominant metastable basins of the system. Given this fixed choice of thresholds, we observed that the number of metastable basins also depends on the persistence length. As an illustration of this, we plot in Fig. \ref{fig:SI-optimal-taup-network} the entropic landscape $-S(h,\mathcal{L})$ with labeled metastable basin locations, Markov state transition graph, and mean transition rates after bootstrapping $\langle K_{ij} \rangle$ for a persistence length near the optimum for cage breaking, $\ell_p=\sigma/2\approx0.89$.

The $\ell_p=\sigma/2$ graph in Fig. \ref{fig:SI-optimal-taup-network}B has more metastable basins than the corresponding graph for $\ell_p=10$ in Fig. \ref{fig4:transition-graphs}B, indicating that the particles cluster in additional ways at this intermediate value of $\ell_p$ that are not significant for larger $\ell_p$. As a specific example, we observed that the equilateral triangle arrangement labeled $h_3$ for the 1D landscape in Fig. \ref{fig2:entropic-landscape}A corresponds to basins 3 and 8 in Fig. \ref{fig:SI-optimal-taup-network}A. At larger $\ell_p$ these basins become insignificant compared to the local entropic landscape and are thus not identified in our $\ell_p=10$ transition graph. Fig. \ref{fig:SI-optimal-taup-network}C also shows that the transition rates are generally larger and more uniform for $\ell_p=\sigma/2$ than the corresponding rates at $\ell_p=10$ from Fig. \ref{fig:SI-bootstrapped_Kij_pi}B, indicating that the system explores the metastable basins more rapidly and evenly near this optimal persistence length.

\subsubsection*{Probability flux in $(|h|, \mathcal{L})$--space}
To complement our coarse-grained Markov state model approach, we calculate the probability distribution and probability flux maps in 2D coarse-grained phase space ($|h|, \mathcal{L}$) of the athermal active Brownian particle dynamics from the main text. We summarize the definitions here, and for a thorough description of the procedure, see the SI of \cite{Battle2016BrokenSystems}.

In short, we must discretize phase space into identical-size square boxes of side length $\Delta x$, each of which will represent a discrete "state" $\alpha$. The probability is simply
\be
p_\alpha = \dfrac{t_\alpha}{t_{\text{total}}},
\ee
where $t_\alpha$ is the total amount of time that the system is in state $\alpha$ over the experiment and $t_{\text{total}}$ is the total duration of the experiment.

The probability flux through state $\alpha$ is defined with respect to the transition rates into and out of $\alpha$ along the two dimensions of phase space. The transition rate from state $\alpha$ to state $\beta$ along the direction $x_i$ is estimated from finite-length time trajectories as
\be
w_{\alpha, \beta}^{(x_i)}=\dfrac{N_{\alpha, \beta}^{(x_i)}-N_{\beta, \alpha}^{(x_i)}}{t_{\text{total}}}.
\ee
Here $N_{\alpha, \beta}^{(x_i)}$ is the number of transitions from state $\alpha$ to state $\beta$ along the direction $x_i$. A generic state will have two neighboring states in each direction (upstream $\alpha^{-}$ and downstream $\alpha^{+}$), resulting in four possible transitions. The probability flux is then defined as
\be \label{eq: prob-flux-definition}
\vec{j}(\vec{x}_\alpha)=\frac{1}{2\Delta x}\begin{pmatrix} w_{\alpha^{-}, \alpha}^{(x_1)} + w_{\alpha, \alpha^{+}}^{(x_1)} \\ w_{\alpha^{-}, \alpha}^{(x_2)} + w_{\alpha, \alpha^{+}}^{(x_2)} \end{pmatrix}.
\ee
States on the boundary of the domain of course have fewer possible transitions. The magnitude of the probability flux defined by Eq. (\ref{eq: prob-flux-definition}) is plotted in the main text for $\ell_p=0.005$ and $\ell_p=10$, and the direction of the probability flux field is illustrated via a streamlines phase plot.

Circulation patterns in the probability flux field over coarse-grained phase space are a signature of broken detailed balance \cite{Battle2016BrokenSystems}. To quantify the statistical significance of an observed circulation pattern in phase space, we compute the contour integral along a particular current loop,

\be
\Omega = \oint \vec{j} \cdot d\vec{l}.
\ee

The contour over which we integrate the probability flux is overlaid on the flux maps in Fig. \ref{fig4:transition-graphs} of the main text. Finite-time sampling even of a steady-state system can lead to spatial correlations in the errors while estimating the current, which could give rise to spurious loops in phase space. To investigate the robustness of current loops, then, we calculate the contour integral $\Omega$ for different bootstrap realizations and test whether the mean is significantly different from zero, as explained in detail in the Markov state model section of the Supplemental Material.

\end{document}

%% file: initialisation.tex
\usepackage{titlesec}
\titleformat{\section}
  {\bf\sffamily}
  {\thesection. }
  {5pt}
  {\MakeUppercase}
\renewcommand{\thesection}{\Roman{section}} 

\titleformat{\subsection}
  {\bf\sffamily}
  {\thesubsection. }
  {5pt}{}
  
\renewcommand{\thesubsection}{\Alph{subsection}}

\usepackage[
 backend=biber,
  citestyle=numeric,
  style=phys,
  biblabel=brackets,
  sorting=none,
  url=false,
  doi=true,
  natbib
]{biblatex}
\usepackage[affil-it]{authblk} 
\usepackage{etoolbox}
\usepackage{lmodern}

\usepackage{eurosym}

\usepackage[utf8]{inputenc}
\usepackage{geometry}
\usepackage[hidelinks]{hyperref}

\usepackage{csquotes}
\usepackage[]{authblk} 
\usepackage{etoolbox}
\usepackage{lmodern}

\usepackage{amsmath}
\usepackage{enumerate}
\usepackage{enumitem}
\usepackage{graphicx}
\usepackage{siunitx}
\usepackage{float}
\usepackage[]{parskip} 

\makeatletter
\patchcmd{\@maketitle}{\LARGE \@title}{\fontsize{16}{19.2}\selectfont\@title}{}{}
\makeatother

\usepackage[normalem]{ulem}

\usepackage{graphicx}
\usepackage{dcolumn}
\usepackage{bm}
\usepackage{tikz}
\usetikzlibrary{math}
\usetikzlibrary{shapes.geometric, arrows}
\usetikzlibrary{positioning}
\usetikzlibrary{shapes,arrows}
\usetikzlibrary{intersections}
\usepackage{csvsimple}
\usepackage{changepage}
\usepackage[tbtags]{mathtools}
\usepackage{siunitx}
\usepackage{csquotes}

\usepackage{float}
\usepackage{subfigure}
\usepackage[
	nonumberlist, 				
	acronym,      				
	nomain,						
	nopostdot,					
]{glossaries}
\usepackage{glossary-superragged}
\usepackage[hidelinks]{hyperref}
\usepackage{color, colortbl}

\usepackage{wrapfig}

\usepackage{pgfplots}
\usepackage{tikz}
\usetikzlibrary{arrows}
\usepackage{amsmath}
\pgfplotsset{compat=newest}
\usepgfplotslibrary{fillbetween}
\usetikzlibrary{calc}
\def\centerarc[#1](#2)(#3:#4:#5)
{ \draw[#1] ($(#2)+({#5*cos(#3)},{#5*sin(#3)})$) arc (#3:#4:#5); }

\usepackage{amssymb}
\let\vec\mathbf
\usepackage{nicefrac}

\usepackage{abstract}

\usepackage{multirow}
\usepackage{tabularx}
\usepackage{booktabs}
\usepackage{array}
\usepackage{longtable}
\newcolumntype{L}[1]{>{\raggedright\let\newline\\\arraybackslash\hspace{0pt}}m{#1}}
\newcolumntype{C}[1]{>{\centering\let\newline\\\arraybackslash\hspace{0pt}}m{#1}}
\newcolumntype{R}[1]{>{\raggedleft\let\newline\\\arraybackslash\hspace{0pt}}m{#1}}

\newacronym{3d}{3D}{three dimensional}
\newacronym{am}{AM}{additive manufacturing}
\newacronym{fdm}{FDM}{fused deposition modeling}
\newacronym{ism}{ISM}{in-space manufacturing}
\newacronym{iss}{ISS}{International Space Station}
\newacronym{fcb}{FCB}{Functional Cargo Block}
\newacronym{dem}{DEM}{discrete element method}
\newacronym{md}{MD}{molecular dynamics}
\newacronym{dc}{DC}{direct-current}
\newacronym[plural=PFCs,firstplural=parabolic flight campaigns (PFCs)]{pfc}{PFC}{Parabolic Flight Campaign}
\newacronym{fft}{FFT}{Fast Fourrier Transform}
\newacronym{cad}{CAD}{Computer Assisted Design}
\newacronym{ptfe}{PTFE}{polytetrafluoroethylene}
\newacronym{ps}{PS}{polystyrene}
\newacronym{nasa}{NASA}{National Aeronautics and Space Administration}
\newacronym{esamm}{ESAMM}{Extended Structure Additive Manufacturing Machine}
\newacronym{amf}{AMF}{Additive Manufacturing Facility}
\newacronym{us}{US}{United States}
\newacronym{usa}{USA}{United States of America}
\newacronym{bmgs}{BMGs}{Bulk Metallic Glasses}
\newacronym{esa}{ESA}{European Space Agency}
\newacronym{si}{SI}{International System of Units, abbreviated from French \textit{Syst\`{e}me International (d'unit\'{e}s)}}
\newacronym{dlr}{DLR}{German Aerospace Center}
\newacronym{liggghts}{LIGGGHTS}{\acrshort{lammps} Improved for General Granular and Granular Heat Transfer Simulations}
\newacronym{lammps}{LAMMPS}{Large-scale Atomic/Molecular Massively Parallel Simulator}
\newacronym{sem}{SEM}{Scanning Electron Microscopy}
\newacronym{RPM}{RPM}{Ramdom Positioning Machine}
\newacronym{rpm}{rpm}{revolutions per minute}
\newacronym{rise}{RISE}{Research Internships in Science and Engineering}
\newacronym{daad}{DAAD}{German Academic Exchange Service, abbreviated from German \textit{Deutscher Akademischer Austauschdienst}}
\newacronym{fsm}{FSM}{finite-state machine}
\newacronym{ir}{IR}{infrared}
\newacronym{pcbs}{PCBs}{Printed Circuit Boards}
\newacronym{pcb}{PCB}{Printed Circuit Board}
\newacronym{mcr}{MCR}{Modular Compact Rheometer}
\newacronym{sff}{SFF}{Solid Freeform Fabrication}
\newacronym{uv}{UV}{ultraviolet}
\newacronym{abs}{ABS}{acrylonitrile butadiene styrene}
\newacronym{hpde}{HPDE}{high density polyethylene}
\newacronym{pei}{PEI}{polyetherimide}
\newacronym{bff}{BFF}{BioFabrication Facility}
\newacronym{lens}{LENS}{Laser Engineered Net Shaping}
\newacronym{cnc}{CNC}{Computer Numerical Control}
\newacronym{ebf3}{EBF$^3$}{Electron Beam Free-Form Fabrication}
\newacronym{leo}{LEO}{Low Earth Orbit}
\newacronym{pc}{PC}{polycarbonate}
\newacronym{crissp}{CRISSP}{Customisable Recyclable International Space Station Packaging}
\newacronym{Athena}{Athena}{Advanced Telescope for High-ENergy Astrophysics}
\newacronym{lbm}{LBM}{Laser Beam Melting}
\newacronym{bam}{BAM}{Federal Institute for Materials Research and Testing, abbreviated from German \textit{Bundesanstalt f\"{u}r Materialforschung und-pr\"{u}fung}}
\newacronym{pbf}{PBF}{powder bed fusion}
\newacronym{eb}{EB}{Electron Beam}
\newacronym{2d}{2D}{two dimensional}
\newacronym{4d}{4D}{four dimensional}
\newacronym{pmma}{PMMA}{polymethylmethacrylate}
\newacronym{1g}{$1g$}{gravity on-ground}
\newacronym{mug}{$\mu g$}{microgravity}
\newacronym{bcm}{BCM}{Box Counting Method}
\newacronym{mct}{MCT}{Mode Coupling Theory}
\newacronym{gmct}{gMCT}{granular Mode Coupling Theory}
\newacronym{itt}{ITT}{Integration Through Transients}
\newacronym{mfc}{MFC}{Mass Flow Controller}
\newacronym{ct}{CT}{computed tomography}
\newacronym{xct}{XCT}{X-ray computed tomography}
\newacronym{cv}{CV}{curriculum vitae}
\newacronym{pi}{PI}{principal investigator}
\newacronym{osp}{OSP}{orthogonal superimposed perturbation}
\newacronym{npi}{NPI}{Network Partnering Initiative}
\newacronym{ecsat}{ECSAT}{European Centre for Space Applications and Telecommunications}
\newacronym{eac}{EAC}{European Astronaut Centre}
\newacronym{estec}{ESTEC}{European Space Research and Technology Centre}
\newacronym{fps}{fps}{frames per second}
\newacronym{pdf}{pdf}{probability density function}
\newacronym{ss}{\textit{SS}}{\textit{Smooth Surface}}
\newacronym{rs}{\textit{RS}}{\textit{Rough Surface}}
\newacronym{rcp}{rcp}{random close packing}
\newacronym{iop}{IoP UvA}{Institute of Physics of the University of Amsterdam}
\newacronym{mp}{MP}{Institute of Material Physics for Space}
\newacronym{elgra}{ELGRA}{European Low Gravity Research Association}
\newacronym{zarm}{ZARM}{Center of Applied Space Technology and Microgravity}
\newacronym{piv}{PIV}{particle image velocimetry}

%% file: colors_define.tex
\definecolor{brickred}{rgb}{0.8, 0.25, 0.33}
\definecolor{darkorange}{rgb}{1.0, 0.55, 0.0}
\definecolor{persiangreen}{rgb}{0.0, 0.65, 0.58}
\definecolor{persianindigo}{rgb}{0.2, 0.07, 0.48}
\definecolor{cadet}{rgb}{0.33, 0.41, 0.47}
\definecolor{turquoisegreen}{rgb}{0.63, 0.84, 0.71}
\definecolor{sandybrown}{rgb}{0.96, 0.64, 0.38}
\definecolor{blueblue}{rgb}{0.0, 0.2, 0.6}
\definecolor{ballblue}{rgb}{0.13, 0.67, 0.8}
\definecolor{greengreen}{rgb}{0.0, 0.5, 0.0}